\documentclass[12pt]{article}
\usepackage[latin1]{inputenc}
\usepackage{latexsym}
\usepackage{graphicx}
\usepackage{epsfig}

\def\.{\cdot}

\def\o{\over}
\def\v{\vec}
\def\a{\alpha}

\def\c{\gamma}
\def\b{\beta}
\def\d{\delta}

\def\P{\Psi}

\def\ep{\epsilon}

\def\p{\partial}

\def\+{\bigoplus}

\def\({\left(}
\def\){\right)}
\def\[{\left[}
\def\]{\right]}
\def\l.{\left.}
\def\r.{\right.}

\def\be{\begin{equation}}
\def\ee{\end{equation}}
\def\bea{\begin{eqnarray}}
\def\eea{\end{eqnarray}}
\def\nn{\nonumber \\ &&}

\def\ber{\begin{array}}
\def\eer{\end{array}}

\begin{document}

\begin{titlepage}

\title{BRS "Symmetry", prehistory and history }

\author{Carlo M. Becchi, University of Genova and INFN sezione di Genova}

\maketitle
\thispagestyle{empty}

\centerline{\bf Abstract}
Prehistory - Starting from  't Hooft's (1971) we have a short look at Taylor's and  Slavnov's works (1971-72) and at the lectures given by  Rouet and Stora in Lausanne-1973 which determine the transition from pre-history to history.

History - We give a brief account of the main analyses  and results of the BRS collaboration concerning the renormalized gauge theories, in particular the method of the  regularization independent, algebraic renormalization, the algebraic proof of ${\bf S}$-matrix unitarity and that of gauge choice independence of the renormalized physics. We conclude this report with a suggestion   to the crucial question: what could remain of BRS invariance beyond perturbation theory.

\footnote{Talk given at {\it A Special day in honour of Raymond Stora, Annecy, July 8, 2011.}}

\end{titlepage} 

\section{Introduction}

On the occasion of Raymond Stora's 80th birthday  a short look at the history of the discovery of BRS "Symmetry" is certainly in order. In particular this historical analysis is crucial in order to single out Raymond's role in the discovery.

I think that it is fair to say that the so called {\it symmetry} revealed by the BRS work is nowadays a well established tool. One should always remember that BRS's is by no means a symmetry stating a correspondence among  physical configurations satisfying the same evolution laws. One should rather speak of equivalence among non physical configurations associated with the same set of physical states and observables. It turns out that this equivalence is described in terms of {\it BRS cohomology} .  BRS invariance defines the  equivalence criterion among the elements of the mentioned classes and the laws of physics appear as the evolution laws among equivalence classes. 

The discovery of BRS invariance marks in particular the transition between BRS prehistory and history. 
Indeed, as in many cases in Physics, the discovery is the consequence of a long sequence of steps, that we arbitrarily  begin with 't Hooft's 1971 paper \cite{gth}, thus giving for achieved former crucial results such as  Faddeev-Popov's functional measure \cite{fp}. After Faddeev-Popov's  work a general agreement was reached about the form of the Lagrangian and hence of Feynman rules, the remaining items to be discussed being essentially renormalizability, gauge choice independence, locality and unitarity, together with possible generalizations to e.g. gravity. For this reason the first analyses were devoted to the extension to the Yang-Mills theory of the Ward Identity of QED which expresses at the level of Green functions the freedom of the scalar component $\p_\mu A^\mu \equiv \p A$ of the gauge field. This is the remnant of gauge invariance after gauge fixing. 

In order to avoid confusions we call Ward identities the relations among Green functions involving the scalar components of gauge vector fields, while we call {\it symmetry identities} those stating the invariance of  Green functions under certain class of transformations of physical fields and operators  and {\it equivalence identities}, those which refer to  non-physical  field and operator  transformations. In this sense the above mentioned papers by  't Hooft,  Taylor \cite{jct} and  Slavnov \cite{aas} dealt with Ward identities, in particular 't Hooft's identity states that $\p A$ has vanishing connected Green functions with physical operators (mass-shell transverse gluon fields) while  Slavnov-Taylor identities describe more general Green functions of  $\p A $ and  gluon fields.

These achievements were followed by the discovery of the variant Slavnov(-Taylor) identity presented by  Rouet and  Stora (RS) in the lecture notes of the {\it Enseignement du troisi\`eme cycle de la Physique en Suisse Romande} in 1973 \cite{rs}, that we call the RS identity.  RS's is in fact not a Ward identity,  indeed its interpretation  is the equivalence  of the Green functions related by a transformation  of the unphysical field variables. RS identity was and still is mistaken as an equivalent version of the Slavnov-Taylor identity. It took a while before the real content appeared clearly marking the transition between prehistory and history of BRS construction. In the next section we shall discuss the physical motivations for the introduction of the new identity.
 
 We shall further discuss the consequences of the new identity in the light of  its  new interpretation. One should mention that this led, first of all, to a drastic simplification of the already presented arguments for unitarity, renormalizabilty and gauge fixing independence of gauge theories. A clear evidence of this  simplification was given by J. Zinn-Justin in his lectures given at Bonn in 1974 \cite{jzj}. At the same time BRS invariance allowed a rigorous  and very general definition of their physical content.

After a short discussion of 't Hooft's and Slavnov-Taylor Ward identities we discuss the origin of the RS identity. Considering this identity from the point of view of BRS invariance we shortly recall its role in the construction of a renormalized theory without insisting on the brute force algebraic method. We just consider the general lines of the renormalization procedure. Then we review the proof of ${\bf S}$-matrix unitarity using the formulation due to Kugo and Ojima which greatly clarifies the relation between unitarity and BRS invariance via the existence of a conserved charge $Q$ in the asymptotic state space. The third main point we are going to discuss is gauge fixing independence that we analyse in the tree approximation leaving the technicalities of the extension to the fully renormalized theory in the original papers.

We conclude this report trying to give an answer to the crucial question: what remains of BRS invariance beyond perturbation theory?
\eject
\section{The new Ward
Identities} 

 As mentioned above, 't Hooft's derivation of  the Ward identity for  Yang-Mills theory proceeds through the invariance condition of the connected Green Functional generator under a particular class of field dependent gauge transformations whose infinitesimal parameter $\omega(x)$ is constrained by the equation:
\be \p_\mu D^\mu \omega (x)=J(x)\simeq 0\ ,\label{1}\ee where $D$ is the covariant derivative and  $J(x)$  an arbitrary infinitesimal classical {\it source}. Whenever possible,  we hide color indices. In 't Hooft's diagrammatic analysis the connected Green Functional generator appears as a circle in the graphic representation of the formulae. The particular choice of the gauge parameter, which is natural in the light of the construction of a Ward identity,  is analogous to that made by Fradkin and Tyutin \cite{ft} with the aim of proving  directly the gauge fixing independence of Yang-Mills theory. For this reason Fradkin and Tyutin  identified $J$ with $\p A $. 

Furthermore 't Hooft limited his analysis to Green functions  involving only physical fields $\phi^{i}_{ph}$, in particular transverse gluons on the mass-shell, and  $\p  A  $. The reason for this choice is  the simplicity of the resulting Ward identity that can be written for the connected Green functions  in the form \footnote{This identity gives a prescription for finite counter-terms if a "consistent" regularization exists, that is, in the absence of anomalous breaking terms. This is in fact the case since dimensional regularization is consistent. The same comment applies to the Taylor-Slavnov identity.}.
:
\be \langle 0|T(\prod_{j=1}^k \p A(x_j)\prod_{i=1}^n \phi^{i}_{ph})|0\rangle_C=0\ .\label{2}\ee

The validity of Eq.(\ref{2}) is proven by diagrammatic techniques, however these are easily translated into the functional language and the strong  form of the quantum action principle\footnote{This is the principle of stationary action in the presence of field sources and taking into account Jacobian corrections which holds true in the case of a consistent regularization. } noticing  that the source terms associated with the physical fields do not contribute since they are gauge invariant. There are instead contributions from the Faddeev-Popov determinant, which is not gauge invariant,  and from the functional measure since the gauge parameter is field dependent, and hence the gauge transformation produces a non-trivial variation of the Jacobian. However  the sum of these contributions vanishes \cite{ft}. In other words the Faddeev-Popov determinant is not invariant under the gauge transformations constrained by Eq.(\ref{1}) but the  Faddeev-Popov functional measure is.

The restriction to physical fields, however particularly convenient for the proof of the Ward identity and without consequences on the renormalization analysis\footnote{Modulo mass singularities.}, has the drawback of hiding locality since physical charged fields are non-local operators. This induced Taylor and Slavnov to  extend 't Hooft's Ward identity by relaxing the restriction to physical fields and considering generic vector field Green functions.
It is clear that new terms appear in the identity which depend on the vector field source and, in particular vanish for physical values of the sources. Considering the quantum action principle it is also clear that the new terms correspond to the variation of the action source term $\int dx\  j_\mu(x)A^\mu(x)$ under the infinitesimal gauge transformation whose parameter satisfies Eq.(\ref{1}). In order to fix our formulae we consider, without introducing any further label, the invariantly (dimensionally) regularized bare Yang-Mills theory. Therefore we write the (bare) action in the general Feynman gauge with parameter $\a$ in the form:\be {\cal A}=\int dx [-{G_{\mu,\nu} G^{\mu,\nu} \o4}- {(\p A)^2\o2\a}-\p^\mu\bar cD_\mu c]\ ,\label{aca}\ee where the ghost field $c$ is assumed Hermitian while $\bar c$ is anti-Hermitian \cite{brs3}\footnote{Notice that this choice is not uniformly done, see in particular \cite{ass}}. The corresponding renormalized theory and the extension to theories without invariant regularizations will be considered in the light of the BRS algebraic method.
Considering the Green functional generator:
\be  \langle 0|T\(\exp(i\int dx j_\mu(x)A^\mu(x))\)|0\rangle\equiv Z[j]\ ,\label{gen1}\ee  and the infinitesimal gauge transformation with parameter $\omega$ satisfying Eq.(\ref{1})  one finds:
\be \int dx \langle 0|T\([{J(x)\p A(x)\o\a}-j_\mu (x)D^\mu \omega (x)]\exp(i\int dy j_\mu(y)A^\mu(y))\)|0\rangle=0\ .\label{s0}\ee  
Now, going just a little farther than the Authors of \cite{jct}  and \cite{aas}  and introducing the Faddeev-Popov fields $c$ and $\bar c$ explicitly, we have the identity:
\bea&&\langle 0|T\(\omega (x)\exp(i\int dz j_\mu(z)A^\mu(z))\)|0\rangle\nn=  -i\int dy \langle 0|T\(c(x)\bar c(y)J(y)exp(i\int dz j_\mu(z)A^\mu(z))\)|0\rangle\nn\equiv  \int dy \langle 0|T\(M^{-1}(x,y)J(y)exp(i\int dz j_\mu(z)A^\mu(z))\)|0\rangle\ .\label{s1}\eea
As a matter of fact the non-local operator $M^{-1}$ was introduced by Slavnov who noticed the relation between this operator and "the Green function of the $c$-particle." 

Combining Eq.s (\ref{s0}) and (\ref{s1}) and selecting the coefficient of the linear term in $J(x)$
one has:
\bea&& \langle 0|T\({\p A(x)\o\a}\exp(i\int dy j_\mu(y)A^\mu(y))\)|0\rangle\nn=i \int dy \langle0|T\(\bar c(x)j_\mu(y)D^\mu c(y)exp(i\int dz j_\mu(z)A^\mu(z))\)|0\rangle\ .\label{s2}\eea This equation, although not appearing in references \cite{jct} and \cite{aas}, is an obvious explicit form of the Slavnov-Taylor identity (see \cite{ass}). The apparent non locality of its right-hand side induced Raymond's group to  the search for a further development. They were convinced that the non local form of Eq.(\ref{s2}) is the direct consequence of not considering the Faddeev-Popov fields as full right, however non-physical, fields. Indeed  there is no source for them in Slavnov's and Taylor's Green functional generator. Therefore Raymond and Alain Rouet introduced the complete "local" Green functional generator:
\be\langle0|(T\exp(i{\cal A}_S))|0\rangle\equiv  \langle 0|T\(\exp(i\int dx[ j_\mu(x)A^\mu(x)+\bar\xi(x)c(x)+\xi(x)\bar c(x)])\)|0\rangle\equiv Z_L[j]\ ,\label{gen2}\ee and considered the quantum action principle identity corresponding to an infinitesimal gauge transformation with parameter $c$.
Here and in the following we adopt the standard notation for the ghost fields, $c$ is the geometrical ghost and $\bar c$ is the multiplier.\footnote {This notation was used by Slavnov in Eq.(15) of his paper \cite{aas}, however there was an exchange of notation in Eq.s (38) and (39) which was followed by RS and then BRS.}

The Rouet and  Stora  (RS) point of view  is essentially different from that of 't Hooft, Slavnov and Taylor, since now the field transformation is local \footnote {We shall see in a moment that this allows a regularization independent renormalizability proof.}. However the compensation between the variation of the Faddeev-Popov determinant and that of the Jacobian of the vector field measure is now absent since, $c$ being an independent  field, the contribution from  the Jacobian of the functional measure vanishes while the  Faddeev-Popov determinant remains not invariant. As a matter of fact the crucial point of the new analysis is the evaluation of  the infinitesimal variation of the Faddeev-Popov determinant, this was done with the acknowledged help of C. Itzykson.

In order to discuss this point we have to use a more explicit notation for the gauge indices which is easily done since here all the fields and sources correspond to elements of the Lie algebra of the rigid gauge group. We denote the ghost field components  by $c^\a(x)$ and we introduce the functional operator $T_\a(x)$ generating an infinitesimal local gauge transformation  whose sole non-trivial action is given by:\be(\Lambda T)A^\a_\mu(x) \equiv \int dy\Lambda^\b(y)T_\b(y)A^\a_\mu(x)=D_\mu\Lambda ^\a(x)\ ,\ee and which satisfies the Lie algebra relation:\be \[T_\a(x) , T_\b(y)\]=\d(x-y)f_{\a , \b}^\c T_\c(x)\ .\label{lie}\ee
From Eq.(\ref{lie}) one finds the identity\footnote{Here we use the convention that the generators $T$ are anti-Hermitian and the structure constants are real and antisymmetric and scale with the coupling constant.}:
\be(cT)^2={1\o2}\int dx c^\a(x)c^\b(x)f_{\a , \b}^\c T_\c(x)\equiv ({c\wedge c\o2}T)\ .\ee
Now it is easy to compute the result of applying the  $(cT)$ operator on  the Faddeev-Popov term in the action which  is  given by ${\cal A}_{\Phi\Pi}=-\int dx \p^\mu \bar c (cT)A_\mu(x) \ . $ Indeed, acting on ${\cal A}_{\Phi\Pi}$ by $(cT)$, after integration by parts, one gets: 
\bea (cT){\cal A}_{\Phi\Pi} &&=\int dx\ \p^\mu \bar c(x) (cT)^2A_\mu(x) =\int dx\ \p^\mu \bar c (x)D_\mu{c\wedge c\o2}(x)\nn =-\int dx\ D_\mu\p^\mu \bar c (x){c\wedge c\o2}(x)=-\int dx \  \bar \xi(x){c\wedge c\o2}(x)\ ,\eea where the last identity follows from the $\bar c$ field equation.

The bare form of the quantum action identity  is found noticing that, with the chosen gauge fixing,  ${\cal A}_{GF}=-1/(2\a)\int dx (\p A)^2  $   one has, using the $c$ field equation:\be 
(cT){\cal A}_{GF} =-{1\o\a}\int dx\ \p A(x)  \p D c(x)=-{1\o\a}\int dx\ \xi(x) \p A(x) \ .\ee
One also has for the source terms :
\be
(cT)\int dx \ j^\mu (x)A_\mu (x) =\int dx \ j^\mu (x)D_\mu c (x)\ .\ee
Combining all these results together one finds, in the tree approximation:
\bea&&\int dx  \langle 0|T\left( [ \ j^\mu (x)D_\mu c (x)- \xi(x) {\p^\mu A_\mu(x)\o\a} -\bar \xi(x){c\wedge c\o2}(x)]\right.\nonumber\\ &&\left.\exp(i\int dy[ j_\mu(y)A^\mu(y)+\bar\xi(y)c(y)+\xi(y)\bar c(y)])\right)|0\rangle=0\label{rs}\eea
This is the general Feynman gauge version of the RS identity which appears in Lausanne Lectures 1973 \cite{rs}. It is a tree approximation identity and contains two non-trivial composite operators for which suitable renormalization prescriptions are, in principle,  needed. The renormalization of the theory together with that of the mentioned composite operators and the general consequences of the RS identity has been the subject of BRS analysis.
 
\eject
\section{The  BRS approach.}

The first important point made by BRS has already been mentioned in the introduction to the present note. RS identity, however inspired by a Ward identity, is an equivalence relation corresponding to the invariance of the quantum action under {\it local} field transformations with Grassmannian parameter $\epsilon$. In the tree approximation these  transformations are given by:
\be A^\mu\to A^\mu+i\epsilon D^\mu c\quad\ ,\quad c\to c+i\epsilon{c\wedge c\o2}\quad\ ,\quad \bar c\to \bar c+i\epsilon {\p^\mu A_\mu\o\a}\ .\label{brs1}\ee These relations  are trivially extended to theories with matter fields and to general linear gauge fixings, i. e. when $\p A$ is replaced by a linear combination of (scalar) fields. The above transformations are known under the name of BRST transformations, and it is commonly agreed that T stays for Tyutin whose contribution to the research in this field is certainly relevant. 

A wide variety of further extensions of the transformation rules is discussed in the literature. Writing Eq.(\ref{brs1}) in the form:
\be \Phi\to\Phi+i\epsilon s\Phi\ ,\label{s}\ee where $\Phi$ is any quantum field, it appears almost immediately from Eq.(\ref{brs1}) that the  $s$ operator is mass-shell nilpotent, i. e. its square vanishes modulo field equations. Indeed this fact is easily verified since the action of $s^2$ is null on all the fields with the exception of $\bar c$ and $s^2\bar c=\p Dc$ which vanishes on the ghost mass-shell. Nilpotency of  $s$ turns out to be crucial for a regularization independent algebraic renormalization,  in the proof of perturbative  ${\bf S}$-matrix unitarity and of its independence of the gauge fixing procedure.  As a matter of fact the mass-shell nilpotency can be transformed into  strict nilpotency introducing the Lautrup-Nakanishi multiplier field $b$ \cite{nlf}. This, although introducing a new unphysical field, greatly simplifies the analysis. We shall briefly discuss  this extension at the end of the present section.

The major part of BRS analysis  \cite{brs1}, \cite{brs2}, \cite{brs3}  is limited to Higgs models in order to avoid  infra-red problems. Since we are not presenting a complete study of gauge theories but we limit ourselves to sketch the main ideas and results, we concentrate on Yang-Mills theories forgetting infra-red problems and discussing e.g. unitarity as if the theory were massive. We present our apologies for this choice.

Technically, in order to profit of BRS invariance in the renormalized theory one has to deal with the composite operators $Dc$ and $c\wedge c$, which, of course, are non-trivially renormalized. In our framework the definition of the suitable renormalization prescriptions requires the introduction of an external non dynamical  field (a source) for every composite operator, hence  in the Yang-Mills case we introduce the sources $\zeta(x)$ for the composite operator $(c\wedge c)/2$  and $\c_\mu(x)$ for $D^\mu c$, adding to the source term of the action, defined in Eq.(\ref{gen2}),  the further term:\be{\cal A}'_S= \int dx [\zeta_\a(x)f^\a_{\b,\c}{c^\b(x)c^\c(x)\o2}+\c_\mu^\a(x)(D^\mu c)_\a(x)]\ .\label{gen3}\ee The new Green functional generator
${\bf Z}\equiv<0|(T\exp(i({\cal A}_S+{\cal A}'_S)))|0>$ satisfies, in the tree approximation, the functional differential equation:
\be\int dx[j_\mu{\d\o\d\c_\mu}-{\xi\o\a}\p_\mu{\d\o\d j_\mu}-\bar\xi{\d\o\d\zeta}](x){\bf Z}\equiv{\cal S}{\bf Z}=0\ .\label{brs2}\ee
Beyond  the normalization conditions of the physical fields and coupling and once the Feynman gauge parameter $\a$ is fixed, a further renormalization constraint is needed in order to fix the ghost renormalization constant and (in the Higgs model case) to constrain further  gauge fixing parameters. This constraint is given by the ghost field equation which in the present case can be written in the form:
\be\xi(x){\bf Z}=-i\p_\mu {\d\o\d\c_\mu(x)}{\bf Z}\ .\label{brs3}\ee This second identity is easily translated in terms of the {\it Effective Action} of the theory which is identified with the functional generator of the 1-particle irreducible Green functions and hence, in the tree approximation, with the classical action. It implies that the effective action depends on the ghost field $\bar c$ and on the external field $\c$ through the linear combination $\c_\mu -\p_\mu \bar c\ .$

Identities (\ref{brs2}) and (\ref{brs3}) allow extending 't Hooft's analysis to a rigorous proof of the regularization independent renormalizability of gauge models. The starting point of the proof is  a general theorem that we call {\it Perturbative Renormalized Quantum Action Principle} \cite{qap}. It asserts that if any identity, corresponding in the tree approximation to the invariance of the action under local field transformations, is broken at a certain loop order  by radiative corrections renormalized  consistently with power counting, this is due to the appearance of breaking terms corresponding to local operators of bounded dimension depending on fields and sources  \footnote{ This is the weak form of the quantum action principle. Let us stress once more that a local identity corresponds to a local field transformation, therefore this theorem does not apply to the above discussed Ward identities.}.

The local identity breaking operators are further constrained by a system of consistency conditions, analogous to the Wess-Zumino \cite{wz} condition for the axial anomaly. Renormalizability is proved by  finding the general solution to the consistency conditions and hence the general form of the breaking and showing  that  breaking terms  of this form can be recursively reabsorbed  introducing suitable finite  Lagrangian counter terms. 

The consistency conditions are obtained by translating Eq.s (\ref{brs2}) and (\ref{brs3}) into a functional differential equations for the effective action:\be\Gamma[A,c,\c-\p \bar c,\zeta]\equiv \bar \Gamma[A,c,\c-\p \bar c,\zeta]-\int dx{(\p A)^2\o2\a}\ ,\ee getting:\be\int dx\[{\d\bar\Gamma\o\d A_\mu}{\d\bar\Gamma\o\d \c^\mu}+{\d\bar\Gamma\o\d \zeta}{\d\bar\Gamma\o\d c}\](x)=0\ .\label{brs5}\ee
A recusive analysis shows that the possible local breaking terms belong to the kernel of a linear nilpotent functional differential operator which is mass-shell equivalent  to: \be\bar s\equiv\int dx\[D_\mu c{\d \o\d A_\mu}+{c\wedge c\o 2}{\d\o\d c}\](x)\ ,\label{sbar}\ee
 and that they are harmless whenever they belong to the image of the same operator. If there were breaking terms not belonging to the image, in mathematical terms they would belong to the cohomology of the same differential operator, in physical terms they would be {\it anomalies.} The above mentioned cohomology is usually called BRS cohomology. 

 In this way identities 
(\ref{brs2}) and (\ref{brs3}) are extended from the tree approximation  to the renormalized level.
The analysis can be extended to the construction of renormalized local physical operators \footnote{The first explicit example of application of the BRS method to a composite physical operator, the gluonic density in QCD, has been given by Kluberg-Stern and Zuber \cite{ksz}.} which in the tree approximation are inserted into the theory coupled to  further external fields (sources) leaving Eq.s (\ref{brs2}) and (\ref{brs3}) invariant. The introduction of the new sources might, in principle, generate new breaking terms whose power counting dimension is strictly related to that of the physical operators, however one can still use the consistency conditions in order to verify the renormalizability of the theory in the presence of the new operators. 

Once the BRS invariant theory is perturbatively renormalized one can  discuss its ${\bf S}$-matrix unitarity \cite{brs1} starting from the coherent state formula \cite{yz}\footnote{We adopt the symmetric definition of 4-dimensional Fourier transform: $(2\pi)^2\tilde f(p)=\int dx\exp(-ipx)f(x)$.}
\be {\bf S}=\Sigma\  {\bf Z}[j]|_{j=0} \equiv :\exp(-\int dp\ \tilde\Phi^a_{in}(p)\Gamma_{a,b}(p){\d\o\d \tilde j_b(p)}) :{\bf Z}[j]|_{j=0}\ ,\label{coh}\ee where we denote by $\Phi^a_{in}$ the asymptotic fields and by $\Gamma_{a,b}$ the corresponding Fourier transformed wave operator. 

The wave operator of the fully interacting theory can be identified with a matrix whose elements are given  by the  two point  1-particle irreducible Green functions. Introducing the effective action $\Gamma [\Phi]$ one has:\be \Gamma_{a,b}(p)\equiv(2\pi)^2 {\d^2\o\d\tilde\Phi_a(p)\d \Phi_b(0)}\Gamma [\Phi]|_{\Phi=0}=-\( (2\pi)^2{\d^2\o\d\tilde j(p)\d j_b(0)}{\bf Z}_c[j]|_{j=0}\)^{-1}_{a,b}\ ,\ee where we have also introduced the connected Green functional ${\bf Z}_c[j]\equiv -i\log({\bf Z}[j])/\hbar$ and, for simplicity,  we exclude fields with non-trivial vacuum expectation value.
The wave operator is constrained, together with the asymptotic fields, by the condition: $\Gamma_{a,b}(p)\langle0|T(\tilde\Phi^b_{in}(p)\Phi_{in}^c(0))|0\rangle=i\d^c_a/(2\pi)^2\ .$ 
 
 The unitarity proof is certainly not the most readable part of BRS production \cite{brs1}\cite{brs2}. It has been highly simplified by Kugo and Ojima \cite{ko} who studied the abelian Higgs model quantized with the Nakanishi-Lautrup multiplier. Once again we sketch the method using the Yang-Mills formulae.

The idea is to single out a suitable  operator $Q$ on the asymptotic scattering state space. This operator must be Hermitian, nilpotent and its kernel must contain only states with non-negative norm. Furthermore $Q$ must satisfy the identity:
\be i[ Q,\Sigma]=[{\cal S}, \Sigma]\ ,\label{q}\ee where ${\cal S}$ is defined in Eq.(\ref{brs2}).
Then one has two main results:
\begin{itemize}
\item The states belonging to the image of $Q$ in the asymptotic scattering state space have zero norm since $Q$ is Hermitian and nilpotent.
\item The ${\bf S}$-matrix transforms the kernel of $Q$ in the asymptotic scattering state space into itself.
\end{itemize}
Indeed from Eq.(\ref{coh}),  Eq.(\ref{q}) and Eq.(\ref{brs2})  one has:
\be  [ Q,{\bf S}]=[Q, \Sigma]{\bf Z}[j]|_{j=0}=-i[{\cal S}, \Sigma]{\bf Z}[j]|_{j=0}=0\ .\label{uni}\ee
One can conclude that, the  ${\bf S}$-matrix being pseudo-unitary (${\bf S}^\dag {\bf S}={\bf I}$) in the indefinite norm asymptotic scattering state space, it is unitary in the kernel of $Q$.

The technical work consists in the identification of the operator $Q$. First of all, starting from the asymptotic wave operators one analyses the asymptotic scattering state space. This is an indefinite metric Fock space generated by the action of the negative frequency Fourier components of the quantized fields on the vacuum state $|0\rangle$ which is annihilated by $Q$. In the Yang-Mills case one gets informations on the wave operators from Eq.(\ref{brs2}) and Eq.(\ref{brs3}). Indeed setting:
\bea \Gamma_{\mu,\nu}(p)=&&(2\pi)^2{\d^2\Gamma\o\d\tilde A^\mu(p)\d A^\nu(0)}\quad\ ,\quad  \Gamma_{\bar c, c}(p)=(2\pi)^2{\d^2\Gamma\o\d\tilde {\bar c}(p)\d c(0)}\ ,\nn
 \Gamma_{\mu, c}(p)=(2\pi)^2{\d^2\Gamma\o\d\tilde\c^\mu(p)\d c(0)}\ ,  \eea one gets: \be \Gamma_{\mu, \nu}(p) \Gamma_{\c_\nu, c}(p)=i{p_\mu\o\a} \Gamma_{\bar c, c}(p)\ ,\ee from Eq.(\ref{brs2}) and: \be ip_\mu  \Gamma_{\nu, c}(p)=\Gamma_{\bar c, c}(p)\ee  from Eq.(\ref{brs3}).
 
  Furthermore selecting in ${\d\o\d\tilde \c_\mu (p)}{\bf Z}$ the term singular at $p^2=0$ one finds:\be {\d\o\d\tilde \c_\mu (p)}{\bf Z}=- \Gamma_{\mu, c}(p){\d\o\d\tilde{\bar\xi }(p)}{\bf Z}+{\bf R}(p)\ ,\label{polep}\ee where ${\bf R}$ has no pole on the ghost mass-shell.
  
  The solutions to the above equations are:\be \Gamma_{\mu,\nu}(p)=\chi(p^2)\[ p_\mu p_\nu - g_{\mu,\nu}p^2\] -\a\  p_\mu p_\nu\quad\ ,\quad \Gamma_{\bar c, c}(p)=p^2\psi(p^2)\quad \ ,\quad \Gamma_{\mu, c}(p)=-ip_\mu\psi(p^2)\label{aawo}\ee Here $\chi(p^2)$ is positive for $p^2<0$,  $\chi(p^2)$ and $\psi(p^2)$ are singular at $p^2=0$ due to infrared effects and have branch cuts for $p^2>0$.  Since $\chi(0_-)$ is directly related to the  norm of  the transverse gluon states we "conclude" that this norm is positive. As already announced following 't Hooft \cite{gth},  we do not consider the difficulty related to the infrared problem, we just regularize our theory replacing everywhere $\psi(p^2)$ and $\chi(p^2)$ by $\psi(p^2-\eta)$ and $\chi(p^2-\eta)$ with  real , positive and "infinitesimal" $\eta$. We notice that the same analysis works without any problem in the massive cases \cite{ko}.

Knowing the wave operators we can study the asymptotic fields. The Fourier transformed vector field is:
\bea\tilde A^\mu_{in}(p) &&=\d(p^2)\left[\Theta(p_0)\(\sum_h\ep_h^\mu(\v p)a_h(\v p)+p^\mu a_l(\v p) +\bar p^\mu a_s(\v p)\)\right.\nn\left.+
\ \Theta(-p_0)\(\sum_h\ep_h^{\mu *}(-\v p)a^\dag_h(-\v p)-p^\mu a^\dag_l(-\v p) -\bar p^\mu a^\dag_s(-\v p)\)\right]\nn +\(1-\a\chi(p^2)\)\d'(p^2)(p\cdot\bar p) p^\mu \(\Theta(p_0) a_s(\v p)-\Theta(-p_0)
a^\dag_s(\v p)\)\label{a}\eea 
where $a_h^\dag$ creates transverse vector particles (gluons) with  helicity $h=\pm 1$, $a_h^\dag$ and $a_h^\dag$ create the scalar and longitudinal gluons respectively, $-\bar p^\mu=(-1)^{\d_{\mu, 0}}  p^\mu$, $\ep_h^\mu(\v p) p_\mu=\ep_h^\mu(\v p)\bar p_\mu=0$ and $\d'$ denotes the derivative of Dirac's measure.

For the ghost fields we have:
\bea&&\tilde c_{in}(p)\ \  =\d(p^2)\left[\Theta(p_0)c(\v p)+
\ \Theta(-p_0)c^\dag(-\v p)\right]\nn
\tilde {(\bar c)}_{in}(p) =\d(p^2)\left[\Theta(p_0)\bar c(\v p)-
\ \Theta(-p_0)\bar c^\dag(-\v p)\right]
\ ,\eea where we have taken into account the Hermiticity properties of the ghost fields. 

Now $Q$ is defined by the following conditions:
\bea&& Q|0\rangle=0\quad\ ,\quad[Q,a_h(\v p)]=[Q,a_s(\v p)]=\{Q,c(\v p)\}=0\nn [Q,a_l(\v p)]=\psi(-\eta)c(\v p)\quad\ ,\quad\{Q,\bar c(\v p)\}=(p\cdot\bar p)\chi(-\eta)a_s(\v p)\ ,\label{brs4}
\eea
from which the $Q$ operator is easily built and it follows that the subspace of the Fock space generated by the creation operators $a^\dag_h\ ,\  a^\dag_s\ ,\ c^\dag $ is the kernel of $Q$. The elements of this kernel with either "c-particles", or scalar gluons, or both kind of particles lie in the image of $Q$ and hence have zero norm.

It remains to verify Eq.(\ref{q}). This is equivalent to the following equation:
\bea&&i\  \left[Q,\int dp\  \(\tilde A^\mu_{in}(p)\Gamma_{\mu,\nu}(p){\d\o\d \tilde j_\nu(p)}-\tilde c_{in}(p)\Gamma_{\bar c, c}(p){\d\o\d \tilde \xi(p)}+\tilde{\bar c}_{in}(p)\Gamma_{\bar c, c}(p){\d\o\d \tilde{ \bar \xi}(p)}\)\right]\nn= \left[{\cal S},\int dp\  \(\tilde A^\mu_{in}(p)\Gamma_{\mu,\nu}(p){\d\o\d \tilde j_\nu(p)}-\tilde c_{in}(p)\Gamma_{\bar c, c}(p){\d\o\d \tilde \xi(p)}+\tilde{\bar c}_{in}(p)\Gamma_{\bar c, c}(p){\d\o\d \tilde{ \bar \xi}(p)}\)\right]\ ,\eea
which, taking into account Eq.(\ref{brs2}) and Eq.(\ref{brs4}) can be written as:
\bea&&i\int dp\  \(p^\mu\tilde c_{in}(p)\psi(p^2)\Gamma_{\mu,\nu}(p){\d\o\d \tilde j_\nu(p)}+{p_\mu\tilde A^\mu_{in}(p)\o\a}\Gamma_{\bar c, c}(p){\d\o\d \tilde {\bar \xi}(p)}\)\nn =-\int dp\  \(\tilde A^\mu_{in}(p)\Gamma_{\mu,\nu}(p){\d\o\d \tilde\c_\nu(p)}- {i\o\a} \tilde c_{in}(p)\Gamma_{ \bar c, c}(p)p_\nu{\d\o\d \tilde j_\nu(p)}\)\ .\eea
This identity is easily verified using Eq.(\ref{polep}) and Eq.(\ref{aawo}). 

Thus, in conclusion we have verified that the operator $Q$ with the desired properties exists and hence  that the ${\bf S}$-matrix is unitary in the physical state space if this is identified with the linear span of the $Q$-equivalent classes of the elements of the kernel of $Q$ provided we consider as $Q$-equivalent two elements whose difference belong to the image of $Q$.

Beyond the definition of the renormalized theory and local operators and the ${\bf S}$-matrix unitarity, the third main consequence of BRS invariance is the independence of physical amplitudes of the gauge choice. Also with respect to this point the original BRS analysis can be remarkably simplified. 
Indeed the original BRS analysis is based on the operator translation of the parametric equations for the Green functions and its most difficult point is the characterization of the operator corresponding to a gauge parameter derivative.The source of  this difficulty is the explicit dependence of Eq.(\ref{brs2}) on the gauge parameter $\a$, this  is however easily avoided introducing the above mentioned  Lautrup-Nakanishi multiplier field $b$ and replacing the third transformation in Eq.(\ref{brs1}) by $\bar c\to \bar c-i \epsilon b$. 

The introduction of the multiplier $b$ implies modifying the action in Eq.(\ref{aca}) into :\be {\cal A}_{ln}=\int dx [-{G_{\mu,\nu} G^{\mu,\nu} \o4}+b\p A+{\a b^2\o2}-\p^\mu\bar cD_\mu c](x)\ ,\label{acb}\ee   adding to the source action in Eq.(\ref{gen2}) the further term ${\cal A}_{ln}=\int dx J(x)b(x)\ , $
 and modifying Eq.(\ref{brs1}) into:
 \be A^\mu\to A^\mu+i\epsilon D^\mu c\quad\ ,\quad c\to c+i\epsilon{c\wedge c\o2}\quad\ ,\quad \bar c\to \bar c-i\epsilon b\ .\label{brs1'}\ee The new BRST transformations define, through Eq.(\ref{s}), a $s_b$ operator which is strictly local and nilpotent\footnote{Notice that  $s$ acting on bosonic Hermitian fields gives fermionic Hermitian operators while, acting upon fermionic Hermitian fields it gives anti-Hermitian bosonic operators.}. The identity in Eq.(\ref{brs2}) is changed into:
 \be\int dx[j_\mu{\d\o\d\c_\mu}+\xi{\d\o\d J}-\bar\xi{\d\o\d\zeta}](x){\bf Z}\equiv{\cal S}_{nl}{\bf Z}=0\ ,\label{brs5}\ee which does not depend explicitly on any gauge fixing parameter and hence constrains all theories with the same gauge field content.
 
 Now the action ${\cal A}_{ln}$ appears as the sum of  its gauge invariant part  that we combine with  the source terms given in Eq.(\ref{gen3}), 
  into: \be\hat {\cal A} = \int dx [-{G_{\mu,\nu} G^{\mu,\nu} \o4}+\c^\mu D_\mu c+\zeta{c\wedge c\o2}](x)\ ,\label{inv}\ee
 and the gauge fixing part which is:
 \be {\cal A}_{gf}= \int dx [b\p A+{\a b^2\o2}-\p^\mu\bar cD_\mu c](x)= -s_b\int dx [\bar c(\p A+\a {b\o2})(x)]\ .\label{gfp}\ee 
 Aiming at an analysis of physically relevant parameters it is worth comparing $\hat {\cal A} +{\cal A}_{gf}$ with the most general renormalizable action compatible with Eq.(\ref{brs5}). For this we translate Eq.(\ref{brs5})  in terms of the effective action getting   the Zinn-Justin \cite{zj2} equation:
\be\int dx\[{\d\Gamma\o\d A_\mu}{\d\Gamma\o\d \c^\mu}+{\d\Gamma\o\d \zeta}{\d\Gamma\o\d c}-b{\d\Gamma\o\d \bar c}\](x)=0\ .\label{zj0}\ee
Restricting this equation to  the tree approximation, we obtain the same equation for the classical action. 
Given an action ${\cal A}$ solution to Eq.(\ref{zj0}) we denote by $\bar{\cal A}$ its restriction to $b=\bar c=0$. $\bar{\cal A}$ satisfies the equation \be\int dx\[{\d\bar{\cal A}\o\d \c^\mu}{\d\o\d A_\mu}+{\d\bar{\cal A}\o\d \zeta}{\d\o\d c}\](x)\bar{\cal A}\equiv {\cal D}_{\bar{\cal A}}\bar{\cal A}=0\ ,\label{zj1}\ee which also defines the nilpotent functional differential operator  ${\cal D}_{\bar{\cal A}}$.

 For simplicity we assume invariance under the action of the rigid gauge group (color symmetry). 
 
 The general renormalizable solution to this equation is known and easily verified to be equivalent, up to field and coupling constant multiplicative renormalization, to $\hat{\cal A}$. Thus, without loss of generality we can identify $\bar{\cal A}$ with $\hat{\cal A}$. 
After this identification the operator ${\cal D}_{\bar{\cal A}}$ coincides with $\bar s$ defined in Eq.(\ref{sbar}).
 
Taking into account the  renormalizability constraint and Faddeev-Popov charge neutrality one sees that ${\cal A}-\hat{\cal A}\equiv {\cal F}$ cannot depend on the external fields $\c_\mu$ and $\zeta$ while it must depend upon either $b$ or $\bar c$. Therefore writing Eq.(\ref{zj0}) in terms of ${\cal F}$ we have:
\be  \[\bar s-\int dx (b{\d\o\d\bar c})(x)\]{\cal F}=0\ ,\ee which implies that:\be {\cal F}=\[\bar s-\int dx (b{\d\o\d\bar c})(x)\]\int dx X[A,b,c,\bar c]\ ,\label{gau1}\ee with:\be \int dx X[A,b,c,\bar c](x)=\int dx\[ \bar c(a_1\p A+a_2 b +a_3A^2+a_4\  \bar c\wedge c)\](x)\ ,\label{gau2}\ee and we have denoted by $\bar c A^2$ the contraction $d_{\a,\b,\c}\bar c_\a A^\mu_\b A_{\mu\ \c}$ assuming that an invariant symmetric $d$ tensor exists. Notice that the functional differential operator: $\bar s-\int dx (b{\d\o\d\bar c})(x)$ coincides with  $s_b$ in Eq.(\ref{gfp}) and hence we have \be{\cal F}=s_bX\ .\ee

In conclusion, in the case of a generic local gauge fixing choice, Yang-Mills theory depends on seven parameter and on a renormalization scale. However, since the first two parameters are multiplicative renormalization constants and since the derivative of the classical action with respect to $a_i$ with $i=1,\cdot\cdot, 4$ is given by $s_b\p_{a_i}X$ the physical content of the renormalized Yang-Mills theory  depends, once the energy scale is given, on a single parameter which must be identified by a suitable normalization condition e. g. involving the Callan-Symanzik beta function \cite{gth2}.

Our analysis is limited to the tree approximation, its extension to all orders of perturbation theory presents technical difficulties which are however perfectly under control \footnote{The origin of these difficulties lies in the non-quadratic nature of $X$ which induces a non-linear $b$ field equation and related renormalization problems. A very general analysis of the possible terms breaking Eq.(\ref{zj0}) or, more precisely, of the identity following from Eq.(\ref{zj0})  after $b$ field integration \cite{zj2}, is presented   in \cite{bbh} where the anti-field formalism \cite{bav} is  exploited.}.

This concludes our sketchy presentation of the Yang-Mills version of the BRS analysis of 4-dimensional gauge theories. Analogous analyses have been  applied to a number of  field theories that we cannot discuss in this report.

 \eject

\section{Conclusions}

It should  clearly appear from the former section of this report that the BRS quantization method identifies renormalized physical operators and physical states with elements of some cohomology classes respectively associated with the  $s$ operator defined in Eq.(\ref{sbar}) and with the charge $Q$ in Eq.(\ref{brs4}). However, in spite of the generality of the method, the explicit results are limited, with some very special exceptions, to renormalized perturbation theory and, either  to strictly local operators, or to perturbation theory asymptotic fields.

An example of physically interesting operators which, to my knowledge have yet to be built even in perturbation theory, is given by  the charged fields localized in space-like cones introduced by Buchholz and Fredenhagen in 1982 \cite{buf}. The existence of these fields is verifiable and should be verified in abelian  and non-abelian Higgs models using Wilson operator product expansion and BRS invariance. I think that the perturbative construction of these operators will give further support to the new approach to  the asymptotic state space given in \cite{buf}. In my opinion the former status of the art in which there are physical operators with a crucial role in the definition of the asymptotic state space, which, however, have a precise physical content only in the asymptotic limit is, in a sense, paradoxical.
Even more problematic is, in my opinion, the definition of the charged fields in the case of unbroken gauge invariance. I think that this second problem has no solution in perturbation theory.

A more important point requiring at least a short discussion is the nature of bounds limiting the BRS analysis to perturbation theory.
Disregarding infrared problems I think that the first major difficulty met beyond  perturbation theory is due to Gribov's horizons \cite{grib}. Roughly speaking the Gaussian functional measure underlying perturbation theory is based on the identification of gauge fields with elements of a single unbounded chart of the  atlas encoding the structure of a highly non-trivial manifold \cite{sing}. One should be able to extend the functional integration to all charts and hence to their overlapping regions. This has been done in the trivial case of 2-dimensional topological gravity \cite{bim}. From the topological gravity results one can figure out, at a very formal level, how the functional measure could be chosen in a 4-dimensional gauge theory keeping BRS invariance unbroken \cite{bgi}. The most relevant character of this measure is its lack of locality which is due to the fact that the distance of a field configuration from the nearest Gribov horizon is not a local  functional.
In a sense, referring to the Yang-Mills case discussed in the previous section, one has to broaden the choice of the operator $X$ in Eq's. (\ref{gau1}) and (\ref{gau2}) by considering suitable non-local operators. Recalling that the discovery of BRS invariance was originated from the rigid observance of strict locality, the fact that, in order to keep this invariance beyond perturbation theory, one has to give up strict locality in the gauge fixing action, appears quite deceiving. However the gauge choice independence of the physical functional measure that follows, as we have seen, from BRS invariance should make  this loss of locality harmless.  All this has however still to be verified.

The ${\bf S}$-matrix unitarity problem beyond perturbation theory is even more problematic. As a matter of fact, even if a non-local version of Eq.(\ref{zj0}) holds true in the non-perturbative theory, and hence  one could still deduce relations among unphysical wave operators, the corresponding asymptotic states could fairly well be lost due e.g. to confinement. If, on the contrary, the  non-perturbative theory, e.g. the electro-weak theory \cite{gow}, should be identified with an effective theory whose corrections are expressed in powers of the total energy rather than of $\hbar$,  ${\bf S}$-matrix unitarity should be related to BRS invariance in much the same way as that presented in this report.

In conclusion, if, as assumed in QCD, local gauge degrees of freedom would characterize the short distance structure of the algebra of the observables, it is expected, on account of asymptotic freedom,  that this structure should still be  identified by the BRS cohomology. 

\eject

\end{document}